\def\fr#1#2{\hbox{${#1\over #2}$}}

\def\ni{\noindent}
\def\vs{\vskip.3cm}

\def\+{{(+)}}  \def\-{ {(-)} }   \def\0{ {(0)} }
\def\1{ {(1)} }  \def\2{ {(2)} }

\def\sq{Q\kern-6pt/}
\def\sQ{Q\kern-12pt\nearrow}
\documentclass[11pt,eqs]{article}

\usepackage{latexsym}
\def\be{\begin{equation}}             \def\ee{\end{equation}}
\def\ba{\begin{array}{rcl}}           \def\ea{\end{array}}
\def\beqa{\begin{eqnarray} }          \def\eeqa{\end{eqnarray} }
\def\beqalign{\begin{eqalign}}        \def\eeqalign{\end{eqalign}}
             
\def\bsubeq{\begin{subequations}}     \def\esubeq{\end{subequations}}
\def\bitem{\begin{itemize}}           \def\eitem{\end{itemize}}

\def\DJ{\leavevmode\setbox0=\hbox{D}\kern0pt
 \rlap{\kern.04em\raise.188\ht0\hbox{-}}D}
\def\dj{\leavevmode\setbox0=\hbox{d}\kern0pt
 \rlap{\kern.215em\raise.46\ht0\hbox{-}}d}

\newcommand{\bd}{\begin{displaymath}}
\newcommand{\ed}{\end{displaymath}}
\begin{document}
\title{ Noncommutativity in space-time extended by Liouville field
\thanks{Work supported in part by the Serbian Ministry of Science and
Environmental Protection, under contract No. 141036.}}
\author{B. Nikoli\'c \thanks{e-mail address: bnikolic@phy.bg.ac.yu} and
B. Sazdovi\'c \thanks{e-mail address: sazdovic@phy.bg.ac.yu}\\
       {\it Institute of Physics, 11001 Belgrade, P.O.Box 57, Serbia}}
%\date{}
\maketitle
\begin{abstract}

The world-sheet quantum conformal invariance can be realized in
the presence of the conformal factor $F$ by inclusion of the
Liouville term. In the background with linear dilaton field,
$\Phi(x)=\Phi_0+a_\mu x^\mu$, the field $F$ becomes a new
noncommutative variable. Therefore, it is natural to extend
space-time with a new coordinate, $F$, in order to unify
expressions for noncommutativity parameter $\Theta^{ij}$ of the
$Dp$-brane space-time coordinates $x^i$, with the part $\Theta^i$
connecting noncommutativity between coordinates $x^i$ and $F$. In
this way we solve the problems of $Dp$-brane noncommutativity in a
more elegant way. The technical advantage is in the fact that in
the extended space-time the action with dilaton field can be
rewritten in dilaton free form.

We use canonical method and extend its application to the
derivation of boundary conditions. From requirement that
Hamiltonian, as the time translation generator, has well defined
derivatives in the coordinates and momenta, we obtain boundary
conditions directly in the canonical form.

\end{abstract}
\vs

\ni {\it PACS number(s)\/}: 02.40.Gh, 11.25.-w, 04.20.Fy.  \par
\section{Introduction}

In Refs.\cite{BNBS1, BNBSL} we explained new possibility in order
to realize world-sheet quantum conformal invariance. Instead of
standard requirement for vanishing $\beta$-functions corresponding
to all background fields (metric, antisymmetric tensor, and
dilaton field), $\beta^G_{\mu\nu}=\beta^B_{\mu\nu}=\beta^\Phi=0$,
we used the fact established in \cite{FCB} that condition
$\beta^G_{\mu\nu}=0$ implies that the third one is constant,
$\beta^\Phi=c$. This constant contribution to the conformal
anomaly has been cancelled by adding Lioville term. Then the
theory depends on arbitrary central charge $c$, and the conformal
invariance is realized in the presence of the conformal factor of
the world-sheet metric, $F$. The Liouville action brings dynamics
to the field $F$. So, in open string theory we should choose its
boundary condition and investigate its contribution to standard
results of noncommutativity without dilaton field
\cite{CDS,CDS1,SW}. It is shown in Refs.\cite{BNBS1,BNBSL,BS2}
that in the presence of linear dilaton field, $\Phi(x)=\Phi_0+a_i
x^i$ (which initially has been investigated in Ref.\cite{LRR}),
the field $F$ is a new noncommutative dynamical variable, while
$Dp$-brane coordinate $x_c=a_i x^i$ is a commutative one.

In order to unify the expressions for noncommutativity parameters
of Refs.\cite{BNBS1,BNBSL,BS2} we are going to treat the conformal
part of the world sheet metric $F$ as an additional coordinate of
some extended space-time with the coordinates $y^M=(x^\mu\, ,F)$
and the metric $G_{MN}$ depending on ordinary metric $G_{\mu\nu}$,
dilaton gradient $a_\mu=\partial_\mu \Phi$ and central charge $c$.
We will investigate the geometry and noncommutativity features of
extended $Dp$-brane, embedded in the extended space-time, and
parameterized by the extended coordinates $y^A=(x^i\, ,F)$,
consisting of $Dp$-brane space-time coordinates $x^i$ and field $F$,
with the corresponding extended metric tensor $G_{AB}$. In the
extended formulation the starting action obtains the dilaton free
form, which simplifies all calculations. Such approach makes the
previous results more transparent and offers their geometrical
interpretation.

We use notation and terminology of Ref.\cite{BNBSL} distinguishing
two descriptions of the same open string theory. In terms of
variable $y^A$ and background field $G_{AB}$, the theory is
described by equations of motion and boundary conditions. The
effective theory, defined on solution of boundary conditions, is
described only by effective equations of motion. This is the
string theory on orbifolds expressed in terms of effective
variables $q^A$ symmetric under transformation $\sigma \to
-\sigma$ and effective background field $G_{AB}^{eff}$. We extend
terminology of Ref.\cite{SW} to the extended theory referring to
metric $G_{AB}$ as \textit{closed string} metric and to
$G_{AB}^{eff}$ as \textit{open string} metric.

As a difference of standard $Dp$-brane metric $G_{ij}$, which is
always regular, the extended metrics can be singular for some
particular relations between background fields. So, we investigate
three cases: (1) both closed and open extended metrics are
regular, (2) extended closed string metric $G_{AB}$ is singular,
and (3) extended open string metric $G_{AB}^{eff}$ is singular.

The singularities of the extended metrics produce the first class
constraints, but with different origins. In the case (2)
singular $G_{AB}$ produce a standard constraint of the
Dirac type. In the case (3) singularity of $G_{AB}^{eff}$ turns some
second class constraints, obtained from the boundary conditions,
into the first class ones. In fact, Poisson brackets of these
constraints close on the extended effective metric $G_{AB}^{eff}$.
The singular directions of this metric project complete set of constraints to the first class
ones.

The first class constraints generate local gauge symmetries. After gauge fixing the first class
constraints and gauge conditions can be treated as second class
constraints. Solving these constraints together with the remaining second
class constraints, we obtain
effective theory.

There is also an important methodological improvement. In subsection 3.2 we
obtain the boundary conditions purely canonically, from the requirement that
Hamiltonian is differentiable in its canonical variables. This is more natural
approach because we intend to treat these conditions as canonical constraints.
The equations of motion for the momenta turn canonical boundary conditions to the standard Lagrangian
ones.

The last part of the article contains concluding remarks and two appendices. The
Appendix A is devoted to the closed and open
string metrics of the extended space-time, and the corresponding zero central charge limit
($c=0$). In the Appendix B we introduce the projectors which
help us to express the results clearly.

\section{Extended $Dp$-brane in extended space-time}
\setcounter{equation}{0}

The action which describes dynamics of the open string in the
presence of the space-time metric $G_{\mu\nu}(x)$, Kalb-Ramond
antisymmetric field $B_{\mu\nu}(x)$, and dilaton scalar field
$\Phi(x)$ is of the form \cite{P}
\begin{equation}\label{eq:action2}
S_{(G+B+\Phi)} = \kappa  \int_\Sigma  d^2 \xi  \sqrt{-g}  \left\{  \left[  {1
\over 2}g^{\alpha\beta}G_{\mu\nu}(x) +{\varepsilon^{\alpha\beta}
\over \sqrt{-g}}  B_{\mu\nu}(x) \right] \partial_\alpha x^\mu
\partial_\beta x^\nu +  \Phi (x) R^{(2)}  \right\} \,  .
\end{equation}
We integrate in the action over the world-sheet surface $\Sigma$
parameterized by $\xi^\alpha=(\tau\, ,\sigma)$ [$(\alpha=0\, ,1)$,
$\sigma\in(0\, ,\pi)$], while the $D$-dimensional space-time is
spanned by the coordinates $x^\mu$ ($\mu=0,1,2,\dots,D-1$). We
denote intrinsic world sheet metric with $g_{\alpha\beta}$, and the corresponding scalar curvature with
$R^{(2)}$.

Three $\beta$-functions characterize the conformal anomaly of the
sigma model (\ref{eq:action2})
\begin{equation}\label{eq:betaG}
\beta^G_{\mu \nu} \equiv  R_{\mu \nu} - \frac{1}{4} B_{\mu \rho
\sigma} B_{\nu}{}^{\rho \sigma} +2 D_\mu a_\nu    \,  ,
\end{equation}
\begin{equation}\label{eq:betaB}
\beta^B_{\mu \nu} \equiv  D_\rho B^\rho{}_{\mu \nu} -2 a_\rho
 B^\rho{}_{\mu \nu}  \,  ,
\end{equation}
\begin{equation}\label{eq:betaFi}
\beta^\Phi \equiv 2\pi \kappa{D-26 \over 6}-R - \fr{1}{24} B_{\mu \rho \sigma}
B^{\mu \rho \sigma} -  D_\mu a^\mu + 4 a^2  \,  ,
\end{equation}
where $R_{\mu \nu}$ and $D_\mu$ are Ricci tensor
and covariant derivative with respect to the space-time metric $G_{\mu\nu}$, while
\begin{equation}
B_{\mu \nu \rho}=\partial_\mu B_{\nu\rho}+\partial_\nu
B_{\rho\mu}+\partial_\rho B_{\mu\nu}\, ,\quad a_\mu=\partial_\mu
\Phi\, .
\end{equation}

It is shown in Ref.\cite{FCB}, that vanishing of
$\beta^G_{\mu \nu}$ and $\beta^B_{\mu \nu}$ gives constant value
of the third $\beta$- function, $\beta^\Phi=c$, and
the non-linear sigma model (\ref{eq:action2}) becomes conformal
field theory. Therefore, Virasoro algebra with central charge
$c$ emerges.

From this point our approach differs from the previous one
\cite{BNBS1}. We retain two conditions, $\beta ^G_{\mu\nu}=0$ and
$\beta^B_{\mu\nu}=0$, but in order to cancel the remaining
conformal anomaly we add the corresponding Wess-Zumino term to the
action (\ref{eq:action2}). In this concrete case the role of
Wess-Zumino term takes the Liouville action
\begin{equation}\label{eq:kinclan}
S_{L}=  -\frac{\beta^\Phi}{2(4\pi)^2\kappa}\int_{\Sigma} d^2 \xi \sqrt{-g}
R^{(2)}\frac{1}{\Delta}R^{(2)}\, ,\quad
\Delta=g^{\alpha\beta}\nabla_\alpha\partial_\beta\, ,
\end{equation}
where with $\nabla_\alpha$ we denote the covariant derivative with
respect to the intrinsic metric $g_{\alpha\beta}$. Note, in this
approach we establish quantum conformal invariance even in the
presence of the field $F$. The complete action, in the conformal
gauge $g_{\alpha\beta}=e^{2F}\eta_{\alpha\beta}$,  takes the form
\begin{equation}\label{eq:2deo1}
S=\kappa \int_{\Sigma}d^2 \xi\bigg [\bigg
(\frac{1}{2}\eta^{\alpha\beta}G_{\mu\nu}+\epsilon^{\alpha\beta} B_{\mu\nu}
\bigg)\partial_{\alpha}x^{\mu}\partial_{\beta}x^{\nu}+2\eta^{\alpha\beta}a_\mu \partial_\alpha
x^\mu
\partial_\beta F+\frac{2}{\alpha}\eta^{\alpha\beta}
\partial_{\alpha}F \partial_{\beta}F\bigg]\,  ,
\end{equation}
where we defined the parameter $\alpha$ as
\begin{equation}\label{eq:alfa}
\frac{1}{\alpha}=\frac{\beta^\Phi}{(4\pi\kappa)^2} \, .
\end{equation}

The space-time with coordinates $x^\mu(\xi)$ is decomposed in
$Dp$-brane part spanned by coordinates $x^i(\xi) \,  (i
=0,1,...,p)$, and the orthogonal ones, $x^a(\xi) \,  (a =
p+1,p+2,...,D-1)$, in such a way that $G_{\mu \nu} = 0$, ($\mu =i
\,  ,\nu = a$). Also we choose the case where the fields
$B_{\mu\nu}$ and $a_\mu$ live only on the $Dp$-brane: $B_{\mu \nu}
\to B_{i j}$, $a_\mu \to a_i$.

In order to simplify the calculations and to offer geometrical
unification of noncommutativity parameters, it is useful to introduce an extended
space-time defined by the coordinates
\begin{equation}
y^M=(x^\mu\, ,F)=(y^A\, ,x^a)\, .
\end{equation}
where an extended $Dp$-brane is parameterized by
\begin{equation}\label{eq:yA}
y^A=(x^i\, ,F) \,  , \quad A\in
\{0,1,\ldots,p+1\}\, .
\end{equation}
The part of the action describing the string oscillation
in $x^a$ directions decouples. We will analyze the rest part described by action
\begin{equation}\label{eq:eaction}
S=\kappa \int_\Sigma d^2 \xi \left[
\frac{1}{2}\eta^{\alpha\beta}G_{AB}+\epsilon^{\alpha\beta}B_{AB}\right]\partial_\alpha
y^A \partial_\beta y^B\, ,
\end{equation}
where the corresponding background fields $G_{AB}$ and $B_{AB}$
are defined in Eq.(\ref{eq:nediagk}). Let us stress that the
action in the presence of dilaton field in extended space-time
has the form of dilaton free action.

The action (\ref{eq:eaction}) can be rewritten in the form
\begin{equation}\label{eq:eactiond}
S=\kappa \int_\Sigma d^2 \xi \left[
\frac{1}{2}\eta^{\alpha\beta}\;{}^\star
G_{AB}+\epsilon^{\alpha\beta}\;B_{AB}\right]\partial_\alpha \,{}^\star y^A \partial_\beta
\,{}^\star y^B\, ,
\end{equation}
where the extended metric ${}^\star G_{AB}$ (\ref{eq:diagk}) is
diagonal, and ${}^\star y^A=(x^i\, ,{}^\star F)$ is defined in
Eq.(\ref{eq:diagc}). Because these two forms of metric are
connected by similarity transformation (\ref{eq:slicnost}),
further we will use the same notation for both of them omitting
the mark ($\star$). The first form is useful for zero central
charge limit $c=\beta^\Phi=0$ ($\alpha\to\infty$), where Liouville
field disappears. The advantages of the second form are
diagonalization of the extended metric and manifest separation of
one variable, ${}^\star F$, which simplify the comparison with
results of Ref.\cite{BNBSL}.

All nontrivial features of the model defined in Eqs.(\ref{eq:eaction}) and
(\ref{eq:eactiond}) follow from the fact that the extended metrics
($G_{AB}$ and the corresponding effective one $G_{AB}^{eff}$) are
singular for some specific choices of the background fields. It is
easy to check that for $\mathcal A\equiv \frac{1}{\alpha}-a^2=0$
and $\tilde \mathcal A\equiv \frac{1}{\alpha}-\tilde a^2=0$ we
have $\det G_{AB}=0$ and $\det G_{AB}^{eff}=0$, respectively.

\section{Canonical analysis of open string theory}
\setcounter{equation}{0}

Because boundary conditions will be treated as canonical
constraints, we will derive them in terms of coordinate and
momenta using canonical method. In this section we assume that
extended metric $G_{AB}$ is regular, while particular cases of
singular extended metrics are discussed in Section 4.

\subsection{Canonical Hamiltonian in terms of currents}

The momenta canonically conjugated to the fields $y^A$ are
\begin{equation}\label{eq:kanimp}
 \pi_A=\kappa (G_{AB}\; \dot y^B-2 B_{AB}\; y'^B)\, .
\end{equation}
Using the definition of the canonical Hamiltonian $\mathcal
H_c= \pi_A \; \dot y^A-\mathcal L$, we obtain
\begin{equation}\label{eq:hamilton}
H_c=\int d\sigma \mathcal H_c\, ,\qquad \mathcal
 H_{c}=T_{-}-T_{+}\, , \quad
T_{\pm}=\mp\frac{1}{4\kappa}(G^{-1})^{AB}\;j_{\pm A} \;j_{\pm
B} \, ,
\end{equation}
where the expression for inverse metric $(G^{-1})^{AB}$ is given in Eq.(\ref{eq:ndmi}), and the current is defined as
\begin{equation}\label{eq:zvezdastruje}
j_{\pm A}= \pi_A+2\kappa\; \Pi_{\pm AB}\; y'^B\, .\quad \left(\Pi_{\pm AB}=B_{AB}\pm\frac{1}{2}
G_{AB}
\right)
\end{equation}

From the basic Poisson bracket algebra
\begin{equation}\label{eq:azs}
\left\lbrace  y^A(\tau,\sigma), \pi_B(\tau,\overline\sigma)\right\rbrace=\delta^A{}_B\delta(\sigma-\overline\sigma)\, ,
\end{equation}
the current algebra directly follows
\begin{equation}\label{eq:algcur}
\left\lbrace  j_{\pm A}, j_{\pm B}\right\rbrace=\pm2\kappa\; G_{AB}\delta'\, ,
\end{equation}
while all opposite chirality currents commute and for simplicity we define
$\delta'\equiv \partial_\sigma \delta(\sigma-\overline\sigma)$.
Consequently, the Poisson bracket between canonical Hamiltonian
and the current $j_{\pm A}$
is proportional to its sigma derivative
\begin{equation}\label{eq:hamstruja}
\left\lbrace  H_c, j_{\pm A}\right\rbrace=\mp  j'_{\pm A}\, .
\end{equation}

This canonical analysis is formally equivalent to the analysis
without dilaton field. But, because we work in extended
space-time, it contains dilaton. The components of the currents
and energy-momentum tensor
\begin{equation}
j^A_{\pm}=G^{AB}j_{\pm B}=\left(
\begin{array}{c}
J^i_\pm\\\frac{1}{2}i^F_\pm
\end{array}
\right)\, ,\quad
J^i_\pm=\left( G^{ij}+\frac{a^i a^j}{\mathcal A}\right)j_{\pm j}-\frac{a^i}{2\mathcal A}i^\Phi_\pm\, ,\quad i^F_\pm=-\frac{1}{\mathcal A}(a^i j_{\pm i}-\frac{1}{2}i^\Phi_\pm)\, ,
\end{equation}
\begin{equation}
T_{\pm}=\mp\frac{1}{4\kappa}\left[  G_{ij}J^i_{\pm}J^j_{\pm}+2(a_i J^i_\pm)i^F_\pm+\frac{1}{\alpha}i^F_\pm i^F_\pm\right] \, ,
\end{equation}

where
\begin{equation}
j_{\pm A}=\left(
\begin{array}{c}
j_{\pm i}\\i^\Phi_\pm
\end{array}
\right)=\left(
\begin{array}{c}
\pi_i+2\kappa \Pi_{\pm ij}x'^j\pm 2\kappa a_i F'\\\pi\pm 2\kappa a_i x'^i\pm \frac{4\kappa}{\alpha}F'
\end{array}
\right)\, ,
\end{equation}
for $c=0$ ($\alpha\to\infty$), are in full agreement with the corresponding ones of the Ref.\cite{BS1}.

\subsection{Hamiltonian derivation of the boundary conditions}\label{tridva}

In Refs.\cite{CDS1,BS2} the open string boundary conditions have been introduced in
Lagrangian formalism and rewritten in terms of canonical variables. Because we
intend to treat the open string boundary conditions as canonical constraints, we will
derive them directly in Hamiltonian form.

The Hamiltonian is a generator of the time translation, so it must be
differentiable in coordinates and momenta. Varying
the Hamiltonian (\ref{eq:hamilton}), we obtain
\begin{equation}\label{eq:varh}
\delta H_c= \delta H_c^{(R)}-\gamma_{A}^{(0)}\delta y^{A} \Big |_0^\pi \,  ,
\end{equation}
where index $R$ denotes the regular term of the form
\begin{equation}\label{eq:varhr}
\delta H_c^{(R)} = \int d \sigma (A_A \delta y^A + B^A \delta \pi_A) \,  ,
\end{equation}
and
\begin{equation}\label{eq:gruslov}
\gamma_{A}^{(0)}=(\Pi_{-}G^{-1})_A{}^B j_{+ B}+ (\Pi_{+}G^{-1})_A{}^B j_{- B}  \,  .
\end{equation}

The Hamiltonian is properly defined canonical variable, when the
boundary term $\gamma_A^{(0)}\delta y^A\Big |_0^\pi$ in equation
(\ref{eq:varh}) vanishes. That is automatically fulfilled for
closed strings, because they do not have endpoints. Assuming that
the variations $\delta y^{A}$ are arbitrary at the open string
endpoints, we obtain the Neumann boundary conditions in the
canonical form, $\gamma_{A}^{(0)} \Big |_0^\pi=0$. On the other
hand, if we suppose that the string endpoints are fixed, $\delta
y^A \Big |_0^\pi=0$, the boundary conditions are known as
Dirichlet boundary conditions. We choose the Neumann boundary
conditions for variables $y^A$. Note that beside the Neumann
boundary conditions on coordinates $x^i$ it also includes the
Neumann boundary condition on Liouville field $F$.

After
imposing the expressions for momenta obtained on their equations of motion,
the boundary conditions reduce to the Lagrangian ones of
Refs.\cite{CDS1,BS2}
\begin{equation}\label{eq:guslovi}
\gamma_{A}^{(0)} = \kappa (-G_{A B} y^{\prime B}  +2  B_{A B} {\dot y}^B )  \,
.
\end{equation}

Checking the consistency of the constraints, with the help of the
relation (\ref{eq:hamstruja}), we obtain an infinite set of
constraints. Using Taylor expansion, we rewrite all the
constraints at $\sigma=0$ in a more compact $\sigma$-dependent
form
\begin{equation}\label{eq:velikog}
\Gamma_{A}(\sigma)=\sum_{n\geq0}\frac{\sigma^n}{n!}\gamma_A^{(n)}(\sigma=0)=(\Pi_{+} G^{-1})_A{}^B  j_{- B}(\sigma)+(\Pi_{-} G^{-1})_A{}^B  j_{+ B}(-\sigma)\, ,
\end{equation}
where
\begin{equation}
\gamma_A^{(n)}\equiv \left\lbrace H_c\, ,\gamma_A^{(n-1)}\right\rbrace=(\Pi_{+} G^{-1})_A{}^B \partial_\sigma^n j_{- B}+(-1)^n (\Pi_{-} G^{-1})_A{}^B \partial_\sigma^n j_{+ B}\, .
\end{equation}
In the same way, we obtain similar expressions from the
constraints at $\sigma=\pi$. From the fact that the differences of
the corresponding constraints at $\sigma=0$ and $\sigma=\pi$ are
also constraints, we conclude that all positive chirality currents
and, consequently, all variables are $2\pi$ periodic functions.
Because of this periodicity the constraints at $\sigma=\pi$ can be
discarded (for more details see Ref.\cite{BS2}).

We complete the consistency procedure finding the Poisson bracket
\begin{equation}
\left\lbrace H_c\, , \Gamma_A\right\rbrace= \Gamma_A'\, ,
\end{equation}
which means that there are no more constraints in the theory.

\section{First class constraints and gauge symmetries}
\setcounter{equation}{0}

In order to finish canonical analysis we have to classify the
constraints. The nature of the constraints depends on the
(non)singularity of the extended metrics, $G_{AB}$ and
$G_{AB}^{eff}$. It turns out that, for some particular choices of
the background fields, they have vanishing determinants, which
produce the first class constraints. According to the Dirac theory
for constrained systems, the first class constraints generate
gauge symmetries in the theory, which existence enables us to fix
nonphysical degrees of freedom.

\subsection{Case of regular metrics ($\mathcal A\neq0\, ,\tilde \mathcal A\neq0$)}

For $\mathcal A\neq0$ we have
\begin{equation}\label{eq:detG}
\det G_{AB}=4\mathcal A\det G_{ij}\neq0\, ,
\end{equation}
so that we are able to solve all velocities in terms of momenta
from Eq.(\ref{eq:kanimp}). In that case there are no constraints
of Dirac type.

The algebra of the constraints $\Gamma_A$ originating from boundary conditions has a simple matrix form
\begin{equation}\label{eq:dklasa}
\left\lbrace \Gamma_A(\sigma)\, ,\Gamma_B(\overline \sigma) \right\rbrace=-\kappa G^{eff}_{AB}\delta' \, ,
\end{equation}
where $G^{eff}_{AB}$ is defined in Eqs.(\ref{eq:efektivna1}) and (\ref{eq:effmex}).
The determinant of $G_{AB}^{eff}$
\begin{equation}\label{eq:detzvezdaeff}
\det G_{AB}^{eff}=4\frac{\tilde \mathcal A^2}{\mathcal A} \det
G_{ij}^{eff}\, ,
\end{equation}
is regular for $\tilde \mathcal A\neq0$ and $\mathcal A\neq0$, and
all constraints are of the second class. We use the assumption
that the standard metrics are regular, $\det G_{ij}\neq0$ and
$\det G^{eff}_{ij}\neq0$.

\subsection{Singularity of the metric $G_{AB}$ ($\mathcal A=0$)}

For $\mathcal A=0$ the determinant of the metric $G_{AB}$
(\ref{eq:detG}) is equal to zero. This means that extended metric
$G_{AB}$ have one singular direction $n^A$ defined in
Eq.(\ref{eq:singG1}). From (\ref{eq:kanimp}) it is clear, that we
are not able to solve all velocities in terms of momenta, and
consequently there must be a primary constraint. For $\mathcal
A=0$, the current
\begin{equation}\label{eq:j}
j\equiv n^A j_{\pm A}=n^A \pi_A+2\kappa n^A B_{AB} y'^B\, ,
\end{equation}
does not depend on velocities and we conclude that it is a
constraint. From the algebra of currents (\ref{eq:algcur}) we
obtain that $j$ commutes with all $j_{\pm A}$, so it is of the
first class.

The canonical Hamiltonian is of the
form
\begin{equation}\label{eq:hamis}
H_c=\int d\sigma \mathcal H_c\, ,\quad \mathcal H_c=T_--T_+\,
,\quad T_{\pm}=\mp \frac{1}{4\kappa}(g^{-1})^{AB} j_{\pm A} j_{\pm
B}\, ,
\end{equation}
where $(g^{-1})^{AB}$ defined in (\ref{eq:malogmetrika}) is the
inverse of the induced metric in the subspace orthogonal to the
vector $n^A$. In order to examine the consistency of the
constraint $j$ we introduce the total Hamiltonian
\begin{equation}
H_T=H_c+\int d\sigma \lambda(\sigma) j(\sigma)\, ,
\end{equation}
where $\lambda$ is a Lagrange multiplier. From the equation
\begin{equation}
\{H_T, j\}\approx 0\, ,
\end{equation}
we conclude that there are no more constraints in the theory and the
multiplier $\lambda$ remains undetermined. This confirms that $j$
is a constraint of the first class.

The first class constraints generates the local gauge symmetry
transformations of an arbitrary variable $X$
\begin{equation}\label{eq:gtrans}
\delta X = \left\lbrace X\, ,G\right\rbrace\, ,\quad G=\int d\sigma \eta(\sigma) j(\sigma)\, .
\end{equation}
If we apply this to the coordinates $y^A$ and
canonically conjugated momenta $\pi_A$, we get
\begin{equation}
\delta y^A=n^A \eta\, ,\quad \delta \pi_A=2\kappa n^B B_{BA}
\eta'\, .
\end{equation}
From the expression
\begin{equation}\label{eq:y0}
\delta y\equiv \delta (\frac{n_{A} y^A}{n^2})=\eta\, ,\quad n^2=G_{AB} n^A n^B\, ,
\end{equation}
we conclude that $y=0$ is a good gauge condition. Note that
\begin{equation}\label{eq:gaugecondition}
n_A y^A=\alpha \mathcal A a_i x^i\, ,\quad n^2=\alpha \mathcal A a^2\, ,
\end{equation}
are equal to zero for $\mathcal A=0$, but their fraction does not
depend on $\mathcal A$ and, consequently, it is finite.

Using the constraint equation $j=0$ and gauge fixing $y=0$ the boundary conditions take the form
\begin{equation}\label{eq:bc2}
\Gamma_A(\sigma)\to \check \Gamma_A=(\Pi_+ g^{-1})_A{}^B j_{- B}(\sigma)+(\Pi_-
g^{-1})_A{}^B j_{+ B}(-\sigma)\, ,
\end{equation}
with the algebra
\begin{equation}
\{\check \Gamma_A(\sigma)\, ,\check \Gamma_B(\overline\sigma)\}=-\kappa \check G_{AB}^{eff}\delta'\, .
\end{equation}
The extended effective metric has the form
\begin{equation}\label{eq:seffm}
\check G^{eff}_{AB}=\Big(-2[(\Pi_+ g^{-1} \Pi_-)+(\Pi_- g^{-1}
\Pi_+)]_{AB}\Big)\Big |_{\mathcal A=0}=\Big( g_{AB}-4(Bg^{-1}B)_{AB}\Big)\Big
|_{\mathcal A=0}\, ,
\end{equation}
with the concrete expression given in Eq.(\ref{eq:cekeff}).
For $\tilde a^2\neq0$ and $\tilde \mathcal A\neq0$ with the help of the relation
\begin{equation}
\det \check G_{AB}^{eff}=4\alpha \tilde a^2 \tilde \mathcal A\det
G_{ij}^{eff}\, ,
\end{equation}
we conclude
that all constraints originating from boundary conditions and
remaining after gauge fixing are of the second class.

\subsection{Singularity of the metric $G^{eff}_{AB}$ ($\tilde \mathcal A=0$)}

From the expression (\ref{eq:detzvezdaeff}) we conclude that,
for $\tilde \mathcal A=0$ and $\mathcal A\neq0$, $\det G_{AB}^{eff}$ has two zeros at $\tilde \mathcal A=0$, which means
that metric $G_{AB}^{eff}$ has two singular directions $\tilde
n^A_1$ and $\tilde n_2^A$ introduced in Eqs.(\ref{eq:vektilde1}) and
(\ref{eq:vektilde2}). According to the algebra of the constraints
(\ref{eq:dklasa}), for $\tilde \mathcal A=0$ and $\mathcal
A\neq0$, the constraints
\begin{equation}\label{eq:prvaklasaeff}
\Gamma_1=\tilde n^A_1 \Gamma_A\, ,\quad \Gamma_2=\tilde n_2^A
\Gamma_A\, ,
\end{equation}
are of the first class, while the rest ones
\begin{equation}
(\Gamma_T)_A= (\hat P_T)_A{}^B \Gamma_B\, ,
\end{equation}
are of the second class. The projector $(\hat P_T)_A{}^B$,
introduced in Eq.(\ref{eq:hatpt}), projects on the subspace
orthogonal to directions $\tilde n_1^A$ and $\tilde n_2^A$.

The
generator corresponding to the first class constraints $\Gamma_1$ and $\Gamma_2$ is
\begin{equation}
G=\int d\sigma [\eta_1(\sigma) \Gamma_1(\sigma)+\eta_2(\sigma) \Gamma_2(\sigma)]\, ,
\end{equation}
with the same form of gauge transformation as in Eq.(\ref{eq:gtrans}).
Using the identities
\begin{equation}
\tilde n_1^A=2(\tilde n_2 BG^{-1})^A\, ,\quad  \tilde n_2^A=2(\tilde n_1 BG^{-1})^A\, ,
\end{equation}
we obtain that the constraints do not depend on the coordinate $y^A$
\begin{eqnarray}
\Gamma_1&=&\frac{1}{2}(\tilde n_1^A+\tilde n_2^A)\pi_A(\sigma)+\frac{1}{2}(\tilde n_2^A-\tilde n_1^A)\pi_A(-\sigma)\, ,\nonumber \\
\Gamma_2&=&\frac{1}{2}(\tilde n_1^A+\tilde n_2^A)\pi_A(\sigma)+\frac{1}{2}(\tilde n_1^A-\tilde n_2^A)\pi_A(-\sigma)\, .\label{eq:pk}
\end{eqnarray}
Consequently, the gauge transformations of the momenta $\pi_A$ are trivial, and we obtain
\begin{equation}\label{eq:g1}
\delta y^A=\frac{1}{2}(\tilde n_1^A+\tilde n_2^A)\left[ \eta_1(\sigma)+\eta_2(\sigma)\right]+\frac{1}{2}(\tilde n_2^A-\tilde n_1^A)\left[ \eta_1(-\sigma)-\eta_2(-\sigma)\right]\, ,\quad
\delta \pi_A=0\, .
\end{equation}
The
particular gauge transformations
\begin{eqnarray}
\delta y_1&=&\frac{1}{2}\left[ \eta_2(\sigma)+\eta_2(-\sigma)\right] +\frac{1}{2}\left[ \eta_1(\sigma)-\eta_1(-\sigma)\right]\, ,\nonumber \\ \delta y_2&=&\frac{1}{2}\left[ \eta_1(\sigma)+\eta_1(-\sigma)\right] +\frac{1}{2}\left[ \eta_2(\sigma)-\eta_2(-\sigma)\right] \label{eq:komponente} \, ,
\end{eqnarray}
\begin{equation}\label{eq:ipsiloni}
y_a\equiv \frac{\tilde n_{a A}y^A}{\tilde n_a^2}\, ,\quad \tilde n_a^2=G^{eff}_{AB}\tilde n_a^A \tilde n_a^B\;(a=1,2)\, ,
\end{equation}
enable us to choose good gauge conditions
\begin{equation}\label{eq:gejdz}
y_1=0\, ,\qquad y_2=0\, .
\end{equation}
As well as in Eq.(\ref{eq:gaugecondition}) $\tilde n_{a A}y^A$ and
$\tilde n_a^2$ vanish for $\tilde \mathcal A=0$, but the variable
$y_a$ is well defined.

\section{$Dp$-brane features}
\setcounter{equation}{0}

Using the solution of the second class constraints, we obtain that
Poisson brackets of some coordinates are non-zero. The effective
theory, defined on these constraints, describes string symmetric
under $\sigma$-parity, which propagates in new, so called
effective background. In the zero central charge limit, $c=0$, we
get the full agreement with the corresponding results of
Ref.\cite{BNBS1}.

\subsection{Solution of constraint equations}

In the case of regular metrics we solve second class constraints
originating from boundary conditions (\ref{eq:velikog}),
$\Gamma_A=0$. In the cases of singular metrics, the first class
constraints and local gauge symmetry appear. After gauge fixing we
can treat the first class constraints and gauge conditions as
second class constraints. For $\mathcal A=0$ we solve gauge
condition $y=0$, first class constraint $j=0$, and, the second
class constraints $\check \Gamma_A=0$ (\ref{eq:bc2}). In the case
$\tilde \mathcal A=0$, solution of the the first class
constraints, $\Gamma_1=0$ and $\Gamma_2=0$, and the second class
ones, $(\Gamma_T)_A=0$, is equivalent to the solution of all
constraints, $\Gamma_A=0$. Then the complete set of equation
consists of $\Gamma_A=0$ and the gauge conditions, $y_1=0$ and
$y_2=0$, (\ref{eq:gejdz}).

In terms of the $\sigma$-symmetric and antisymmetric parts of coordinates and momenta
\begin{eqnarray}
q^A(\sigma)=\frac{1}{2}\left[y^A (\sigma)+y^A (-\sigma)\right] \,  ,
\quad \overline{q}^{A}(\sigma)=\frac{1}{2}\left[
y^{A}(\sigma)-y^{A}(-\sigma)\right] \, ,\nonumber\\
p_A(\sigma)=\frac{1}{2}\left[\pi_A (\sigma)+ \pi_A (-\sigma)\right] \,  ,
\quad  \overline{p}_{A}(\sigma)=\frac{1}{2}\left[
\pi_{A}(\sigma)-\pi_{A}(-\sigma)\right]\, ,\label{eq:mena1}
\end{eqnarray}
the constraints $\Gamma_{A}(\sigma)$ have the form
\begin{equation}\label{eq:veza1}
\Gamma_A=2(B G^{-1})_A{}^B  p_B-\kappa  G_{AB}^{eff} \overline q'^B+\overline p_A\, .
\end{equation}
Solving the corresponding set of equations, we obtain that the solution in all three considered cases has the same form
\begin{equation}\label{eq:resing}
y^A_{Dp}(\sigma)=Q^A(\sigma)-2\Theta^{AB}\int_0^\sigma d\sigma_1
P_B(\sigma_1)\, ,\quad \pi_A^{Dp}(\sigma)=P_A(\sigma)\, ,
\end{equation}
where
\begin{equation}
y^A_{Dp}=(P_{Dp})^A{}_B y^B\, ,\quad \pi_A^{Dp}=(P_{Dp})_A{}^B \pi_B\, ,\quad Q^A=(P_{Dp})^A{}_B q^B\, ,\quad P_A=(P_{Dp})_A{}^B p_B\, .
\end{equation}
The projectors $(P_{Dp})_A{}^B$ are given in the Table 1, while the concrete expressions for $(\check P_T)_A{}^B$ and $(\hat P_T)_A{}^B$ are defined in Eqs.(\ref{eq:ptcek}) and (\ref{eq:hatpt}), respectively.

In the case (1), equation (\ref{eq:resing}) contains all
components, because $y^A_{Dp}(\sigma)=y^A(\sigma)$, and we
introduce this notation just in order to unify expressions for all
solutions.

\begin{table}[h]
\begin{center}
\begin{tabular}{|c|c|c|c|}\hline
\textbf{Case} & $\tilde \mathcal A\neq0\, ,\mathcal A\neq0$ &
$\mathcal A=0$ & $\tilde \mathcal A=0$\\
\hline\hline $(P_{Dp})_A{}^B$ & $\delta_A{}^B$ & $(\check
P_T)_A{}^B$ & $(\hat P_T)_A{}^B$\\ \hline
\end{tabular}
\caption{Projectors}
\end{center}
\end{table}

In the cases (2) and (3), corresponding to $\mathcal A=0$ and
$\tilde \mathcal A=0$ respectively, the solution (\ref{eq:resing})
does not contain two directions determined by the vectors $\tilde
n_1^A$ and $\tilde n_2^A$. These directions satisfy Dirichlet
boundary conditions, while the corresponding canonically
conjugated momenta are equal to zero
\begin{equation}
y_1\Big |_0^\pi=0\, ,\quad y_2\Big |_0^\pi=0\, ,\quad \pi_1=0\,
,\quad \pi_2=0\, ,
\end{equation}
where $y_a\, ,(a=1\, ,2)$ is introduced in Eq.(\ref{eq:ipsiloni}).

In all cases tensor
\begin{equation}\label{eq:tetka1}
\Theta^{AB}=-\frac{1}{\kappa}(g_{eff}^{-1} B g^{-1}_{Dp} P_{Dp})^{AB}\, ,
\end{equation}
can be written in the same form and it is manifestly
antisymmetric. The corresponding metrics are given in the Table 2,
where the particular expressions for metrics are given in Appendix
A and for projectors in Appendix B.

\begin{table}[t]
\begin{center}
\begin{tabular}{|c|c|c|c|}\hline
\textbf{Case} & $\tilde \mathcal A\neq0\, ,\mathcal A\neq0$ &
$\mathcal A=0$ & $\tilde \mathcal A=0$\\
\hline\hline $(g_{eff}^{-1})^{AB}$ & $(G_{eff}^{-1})^{AB}$ &
$(\check G_{eff}^{-1}\check P_T)^{AB}$ & $(G_{eff}^{-1}\,\hat
P_T)^{AB}$\\ \hline $(g_{Dp}^{-1})^{AB}$ & $(G^{-1})^{AB}$ &
$(g^{-1})^{AB}$ & $(G^{-1})^{AB}$\\ \hline
\end{tabular}
\caption{Extended metrics}
\end{center}
\end{table}

In the case (1), when both closed and open string metric, $G_{AB}$
and $G_{AB}^{eff}$, are regular, the component form of the tensor
$\Theta^{AB}$ is
\begin{equation}\label{eq:tetka2}
\Theta^{AB}=\left(
\begin{array}{cc}
\Theta^{ij} & \Theta^i\\
-\Theta^j & 0
\end{array}
\right) \, ,
\end{equation}
where
\begin{equation}\label{eq:tenzorteta}
\Theta^{ij}=-\frac{1}{\kappa}(G_{eff}^{-1}\check \Pi_T^0
BG^{-1}\check \Pi_T^0)^{ij}\, ,\quad \Theta^i=\frac{1}{2\kappa
\tilde \mathcal A}(B\tilde a)^i\, ,
\end{equation}
and
\begin{equation}
(\check \Pi_T^0)_i{}^j=\delta_i{}^j+\frac{a_i \tilde a^j}{\tilde
\mathcal A} \, .
\end{equation}
In the basis where both ordinary and effective metrics are diagonal, (\ref{eq:tetka2}) can be transformed as
\begin{equation}\label{eq:tetka123}
{}^\star \Theta^{AB}=-\frac{1}{\kappa}({}^\star G_{eff}^{-1}B\;{}^\star G^{-1})^{AB}=\left(
\begin{array}{cc}
\Theta^{ij} & 0\\
0 & 0
\end{array}
\right) \, ,
\end{equation}
with the same expression for $\Theta^{ij}$ as in (\ref{eq:tenzorteta}).

The antisymmetric parameter $\Theta^{AB}$ in the cases (2) and (3)
has the same form in diagonal and non-diagonal representation
\begin{equation}\label{eq:nekom}
\Theta^{AB}=\left(
\begin{array}{cc}
\Theta^{ij} & 0\\
0 & 0
\end{array}\right)\, ,
\end{equation}
where
\begin{equation}
\Theta^{ij}=-\frac{1}{\kappa}(G_{eff}^{-1}\Pi_T B G
^{-1}\Pi_T)^{ij}\, ,\quad (\Pi_T)_i{}^j=\delta_i{}^j-\frac{a_i
\tilde a^j}{\tilde a^2}-\frac{4}{\tilde a^2-a^2}(Ba)_i(\tilde a
B)^j\, .
\end{equation}
The component form of the above results are in full accordance with the results of the Refs.\cite{BNBS1,BNBSL}.

\subsection{Noncommutativity}

From the Poisson brackets of the basic string variables
(\ref{eq:azs}), we calculate the corresponding ones of the
effective variables
\begin{equation}\label{eq:pzagrada}
\{Q^{A}(\tau,\sigma)\, ,P_{B}(\tau,\overline{\sigma})\}=
(P_{Dp})^A{}_B\delta_{s}(\sigma,\overline{\sigma})\, ,
\end{equation}
where $\delta_{s}(\sigma,\overline\sigma)=\frac{1}{2}[\delta(\sigma-\overline\sigma)+\delta(\sigma+\overline\sigma)]$.

Separating the center of mass variables
\begin{equation}\label{eq:cenmassX}
(y_{Dp}^A)_{cm}=\frac{1}{\pi}\int_0^\pi d\sigma y_{Dp}^A(\sigma)\, ,\quad
y_{Dp}^A(\sigma)=(y^A_{Dp})_{cm}+Y_{Dp}^A(\sigma)\, ,
\end{equation}
we obtain
\begin{equation}
\{ Y_{Dp}^A(\sigma), Y^B_{Dp}(\overline \sigma) \}=\Theta^{AB}\Delta(\sigma+\overline
\sigma)\, ,
\end{equation}
where the function $\Delta(x)$ is defined as
\begin{equation}\label{eq:DELTA}
\Delta(x)=\left\{\begin{array}{ll}
-1 & \textrm{if $x=0$}\\
0 & \textrm{if $0<x<2\pi$}\, .\\
1 & \textrm{if $x=2\pi$} \end{array}\right .
\end{equation}

From the identity
\begin{equation}\label{eq:identitet1}
\frac{\alpha}{2}\Theta^{ij}a_j+\Theta^i=0\, ,
\end{equation}
which holds for components defined in Eq.(\ref{eq:tenzorteta}), we conclude that combination $F+\frac{\alpha}{2}a_i x^i={}^\star F$
is a commutative variable. The same result follows directly from the diagonal form of
the noncommutativity parameter (\ref{eq:tetka123}).

In the case (1) where extended metrics are regular there are one
commutative and $p+1$ noncommutative variables. In other two
cases, where either ordinary or effective extended metric are
singular, the coordinates $y_a$ ($a=1\, ,2$) satisfy Dirichlet
boundary conditions and decrease the number of the $Dp$-brane
dimensions. Because the variable ${}^\star F$ is a commutative
one, there are $p-1$ noncommutative variables in the theory. The
expressions for noncommutativity parameter $\Theta^{ij}$ between
coordinates $x^i$, and $\Theta^i$ between $x^i$ and the field $F$,
are unified.

\subsection{Effective theory}

Let us introduce the effective current
\begin{equation}
\tilde j_{\pm A}=P_A\pm\kappa g^{eff}_{AB} Q'^B\, ,\quad
g^{eff}_{AC}(g_{eff}^{-1})^{CB}=(P_{Dp})_A{}^B\, ,
\end{equation}
which live in subspace defined by projector $(P_{Dp})_A{}^B$
playing the role of unity. Using the solution (\ref{eq:resing}),
we correlate it with corresponding one defined in
(\ref{eq:zvezdastruje}) and get
\begin{equation}
j_{\pm A}=\pm2(\Pi_\pm
g_{eff}^{-1})_A{}^B  \tilde j_{\pm B}\, .
\end{equation}
Substituting these relations in the canonical Hamiltonian
[(\ref{eq:hamilton}) or (\ref{eq:hamis})], we obtain an effective
energy-momentum tensor and Hamiltonian
\begin{equation}
T_{\pm}=\mp\frac{1}{4\kappa}(g_{eff}^{-1})^{AB}  \tilde j_{\pm A}  \tilde
j_{\pm B}\equiv \tilde T_{\pm} \, , \quad \mathcal{H}_{c}=\tilde T_--\tilde T_+\equiv\tilde
\mathcal{H}_{c}  \,  .
\end{equation}

The effective theory is defined in the phase space spanned by the
coordinates $Q^A$ and momenta $P_A$. The expressions $\tilde
T_{\pm}$ satisfy Virasoro algebra. Consequently, the effective
theory is a string theory constrained to the subspace defined by
projector $(P_{Dp})_A{}^B$ and symmetric under $\sigma$-parity but
propagating in the effective background $G_{AB}\to g^{eff}_{AB}$,
$B_{AB}\to 0$.

\subsection{Zero central charge limit $c=0$ ($\alpha\to\infty$)}

The expressions in the non-diagonal form are technically more
convenient for applying zero central charge limit $c=0$
($\alpha\to\infty$). All details concerning zero central charge
limit are expressed in Appendix A.2. For $c=0$ we have
\begin{equation}
\pi_A\to{}^0\pi_A=\kappa({}^0 G_{AB}\dot y^B-2B_{AB}y'^B)\, ,\quad
j_{\pm A}\to {}^0 j_{\pm A}= {}^0 \pi_A+2\kappa {}^0\Pi_{\pm
AB}y'^B\, ,
\end{equation}
where ${}^0\Pi_{\pm AB}=B_{AB}\pm\frac{1}{2} \,{}^0 G_{AB}$ and
${}^0 G_{AB}$ is defined in Eq.(\ref{eq:Gnula}). Index ${}^0$ is
chosen to signify the condition $c=\beta^\Phi=0$.

In the case (1), where both extended metrics are regular ($\tilde \mathcal A\to-\tilde a^2\neq0\, ,\mathcal A\to-a^2\neq0$), the noncommutativity parameter takes the form
\begin{equation}\label{eq:tetanula}
{}^0 \Theta^{AB}=\left(
\begin{array}{cc}
{}^0 \Theta^{ij} & {}^0 \Theta^i\\
-{}^0 \Theta^j & 0
\end{array}\right)\, ,
\end{equation}
where
\begin{equation}
{}^0 \Theta^{ij}=-\frac{1}{\kappa}(\tilde P_T B P_T^0)^{ij}\, ,\quad {}^0 \Theta^i=-\frac{(B\tilde a)^i}{2\kappa \tilde a^2}\, ,
\end{equation}
and the projectors $(P_T^0)_i{}^j$ and $\tilde P_T^{ij}$ are defined in Eq.(\ref{eq:tildeptij}).

In cases (2) and (3) in the limit $c=0$ the noncommutativity
parameters have the same form
\begin{equation}\label{eq:sintetanula}
{}^0 \Theta^{AB}=\left(
\begin{array}{cc}
{}^0 \Theta^{ij} & 0\\
0 & 0
\end{array}\right)\, .
\end{equation}
For case (2) of singular ${}^0 G_{AB}$ ($\mathcal A\to-a^2=0$) the
noncommutativity parameter is defined as
\begin{equation}
{}^0 \Theta^{ij}=-\frac{1}{\kappa}(G_{eff}^{-1}P_T^1 B G^{-1} P_T^1)^{ij}\, ,\quad (P_T^1)_i{}^j=\delta_i{}^j-\frac{4}{\tilde a^2}(Ba)_i(\tilde aB)^j\, ,
\end{equation}
while in the case (3) of singular effective extended metric ${}^0 G_{AB}^{eff}$ ($\tilde \mathcal A\to-\tilde a^2=0$) we have
\begin{equation}
{}^0 \Theta^{ij}=-\frac{1}{\kappa}(G_{eff}^{-1}\hat P_T^1 B G^{-1} \hat P_T^1)^{ij}\, ,\quad (\hat P_T^1)_i{}^j=\delta_i{}^j+\frac{4}{a^2}(Ba)_i(\tilde aB)^j\, .
\end{equation}
Note that in the zero central charge limit identity
(\ref{eq:identitet1}) turns to identity $\Theta^{ij}a_j=0$, so
that commutative variable $\frac{1}{\alpha}{}^\star
F=\frac{1}{\alpha}F+\frac{1}{2}a_i x^i$ turns to $a_i x^i$.

All these results are in full correspondence with the expressions
obtained in framework without Liouville term \cite{BNBS1}.

\section{Concluding remarks}
\setcounter{equation}{0}

In this article we considered noncommutativity properties of the
space-time extended by the conformal part of the world-sheet
metric $F$. The field $F$, introduced by Liouville term, allows us
to establish the quantum conformal invariance without using the
dilaton space-time equation of motion, $\beta^\Phi=0$. In fact,
after imposing the space-time equation of motion,
$\beta^G_{\mu\nu}=0$, the Liouville action cancels the remaining
constant anomaly, $\beta^\Phi=c$, and makes the conformal part of
the world sheet metric, $F$, dynamical variable.

So it is natural to consider the extended space-time with the coordinates $y^M=(x^\mu\, ,F)$ and the metric $G_{MN}$. An extended $Dp$-brane, parameterized by extended string coordinates $y^A=(x^i\, ,F)$ with the corresponding closed string metric $G_{AB}$, is emebedded in the extended space-time. In this way in Eq.(\ref{eq:tetka1}) we unified the expressions for noncommutativity parameter $\Theta^{ij}$ between $Dp$-brane space-time coordinates $x^i$ with the noncommutativity parameter $\Theta^i$ between $x^i$ and the field $F$.

When the both extended metrics are regular ($\mathcal A\neq0$, $\tilde
\mathcal A\neq0$) the analysis in the extended formulation is completely equivalent to
the dilaton free case. Applying known results of the dilaton free
case on extended $Dp$-brane, we independently derived the results of
Refs.\cite{BNBS1,BNBSL,BS2}, in a much simpler way. We show that both
noncommutative parameters, $\Theta^{ij}$ and $\Theta^i$, are just components of
the extended noncommutativity parameter. One $Dp$-brane coordinate,
${}^\star F=F+\frac{\alpha}{2}a_i x^i$, is commutative.
Consequently, number of noncommutative variables is the same as in
the absence of dilaton field.

In the case (2) the closed string metric $G_{AB}$ has one singular
direction, while in the case (3) there are two singular directions
of the open string metric $G^{eff}_{AB}$. In both cases the first
class constraints appear in the theory generating local
symmetries. Fixing the gauge, the first class constraints and
gauge conditions behave like second class constraints. Solving the
second class constraints we obtain an effective theory expressed
in terms of the effective variables $Q^A$, symmetric under
$\sigma$-parity transformation, and corresponding effective metric
$g_{AB}^{eff}$. We conclude that $Dp$-brane is described by one
commutative, ${}^\star F$, and $p-1$ noncommutative coordinates.

All cases can be explained from the unique point of view. As a
consequence of the relation between background fields, some
coordinates $y_{a}=\frac{\tilde n_{a A}y^A}{\tilde n_a^2}$ satisfy
Dirichlet boundary conditions, and change the dimensionality of
the $Dp$-brane. The physical $Dp$-branes are defined by projections
$(P_{Dp})^A{}_B$ with dimensions $D_{Dp}=(P_{Dp})^A{}_A$. The
variable ${}^\star F$ is commutative in all three cases, $y_c={}^\star F$, while all
other directions are noncommutative, because $y_{Dp}^A$ depends on
both the coordinates $Q^A$ and momenta $P_A$ [see
Eq.(\ref{eq:resing})]. So, the number of commutative coordinates
$N_c$ is equal to $1$, and the number of noncommutative
coordinates is $N_{nc}=D_{Dp}-1$. This analysis is summarized in the
Table 3.

\begin{table}[h]
\begin{center}
\begin{tabular}{|c|c|c|c|c|c|c|c|}\hline
Case & $y_a$ & $(P_{Dp})_A{}^B$ & $D_{D_p}$ &  $N_{nc}$ & $y_c$ \\ \hline\hline 1. & $-$ &
$\delta_A{}^B$ & p+2 & p+1 & ${}^\star F$ \\ \hline 2. & $y_1\, ,y_2$ &
$(\check P_T)_A{}^B$ & p &  p-1 & ${}^\star F$\\
\hline 3. & $y_1\, ,y_2$ & $(\hat P_T)_A{}^B$ & p & p-1 & ${}^\star F$
\\ \hline
\end{tabular}
\caption{$Dp$-brane features}
\end{center}
\end{table}

In this article we also introduced one methodological improvement
and derived boundary conditions by canonical methods. Demanding
that canonical Hamiltonian as time translation generator is
differentiable in its canonical variables, we obtain the boundary
conditions purely canonically. We treated boundary conditions as
canonical constraints, so this approach seems to be more natural.
The equations of motion for canonical momenta give the standard
Lagrangian form of the boundary condition.

All results of this paper agree with the corresponding ones in
component notation of Refs.\cite{BNBS1,BNBSL,BS2}. The advantage
of the extended space-time approach is to write the action in the
presence of dilaton field in the dilaton free form. In such a way
we unify the expressions for noncommutativity parameters.

\appendix

\section{Extended space-time}
\setcounter{equation}{0}

In this appendix we will introduce the metrics in diagonal and
non-diagonal form, and the similarity transformation which
connects them. Also we give the expressions for metrics in the
zero central charge limit $c=0$ ($\alpha\to\infty$) in order to
compare them with the results of Ref.\cite{BNBS1}.

\subsection{Extended metrics in non-diagonal and diagonal form}

Let us introduce the coordinates of the extended $Dp$-brane $y^A$
and the corresponding background fields $G_{AB}$ and $B_{AB}$
\begin{equation}\label{eq:nediagk}
y^A=\left(
\begin{array}{c}
x^i\\F
\end{array}\right)\, ,\quad
G_{AB}=\left(
\begin{array}{cc}
G_{ij} & 2a_i\\
2a_j & \frac{4}{\alpha}
\end{array}\right)\, ,\quad B_{AB}=\left(
\begin{array}{cc}
B_{ij} & 0\\
0 & 0
\end{array}\right)\, .
\end{equation}
We can diagonalize $G_{AB}$ applying similarity transformation to
vectors
\begin{equation}\label{eq:vektori}
{}^\star V^A=M^A{}_B V^B\, ,\quad {}^\star V_A=[(M^{-1})^T]_A{}^B V_B\, ,
\end{equation}
where
\begin{equation}\label{eq:slicnost}
M^A{}_B=\left(
\begin{array}{cc}
\delta^i{}_j & 0\\
\frac{\alpha a_j}{2} & 1
\end{array}\right)\, ,\quad (M^{-1})^A{}_B=\left(
\begin{array}{cc}
\delta^i{}_j & 0\\
-\frac{\alpha a_j}{2} & 1
\end{array}\right)\, .\quad (\det M=1)
\end{equation}
From transformation laws for vectors (\ref{eq:vektori}) we can derive the corresponding ones for arbitrary tensors.
Marking variables in diagonal form by ($\star$) we obtain
\begin{equation}\label{eq:diagc}
{}^\star y^A=\left(
\begin{array}{c}
x^i\\{}^\star F
\end{array}\right)\, ,\quad {}^\star F=F+\frac{\alpha}{2}a_i x^i\, ,
\end{equation}
and
\begin{equation}\label{eq:diagk}
{}^\star G_{AB}=\left(
\begin{array}{cc}
{}^\star G_{ij} & 0\\
0 & \frac{4}{\alpha}
\end{array}\right)\, ,\quad {}^\star B_{AB}=B_{AB}\, ,\quad {}^\star G_{ij}=G_{ij}-\alpha a_i a_j\, .
\end{equation}

The inverse of the metric $G_{AB}$ and ${}^\star G_{AB}$ are of the form
\begin{equation}\label{eq:ndmi}
(G^{-1})^{AB}=\left(
\begin{array}{cc}
({}^\star G^{-1})^{ij} & -\frac{a^i}{2\mathcal A}\\
-\frac{a^j}{2\mathcal A} & \frac{1}{4\mathcal A}
\end{array}
\right)\, ,\quad ({}^\star G^{-1})^{AB}=\left(
\begin{array}{cc}
({}^\star G^{-1})^{ij} & 0\\
0 & \frac{\alpha}{4}
\end{array}
\right) \, ,
\end{equation}
where
\begin{equation}
({}^\star G^{-1})^{ij}=G^{ij}+\frac{a^i a^j}{\mathcal A}\, ,\quad
\mathcal A\equiv \frac{1}{\alpha}-a^2\, .
\end{equation}
Because metrics $G_{AB}$ and ${}^\star G_{AB}$ are connected by
similarity transformation, their determinants are equal and have a
form
\begin{equation}
\det G_{AB}=\det\; {}^\star G_{AB}=4\mathcal A\det G_{ij}\, .
\end{equation}

The corresponding effective metrics are
\begin{equation}\label{eq:efektivna1}
G^{eff}_{AB}=G_{AB}-4(BG^{-1}B)_{AB}=\left(
\begin{array}{cc}
\tilde G_{ij} & 2a_i\\
2a_j & \frac{4}{\alpha}
\end{array}
\right)\, ,
\end{equation}
\begin{equation}\label{eq:effmex}
{}^\star G^{eff}_{AB}={}^\star G_{AB}-4(B\;{}^\star G^{-1}B)_{AB}=\left(
\begin{array}{cc}
{}^\star G^{eff}_{ij} & 0\\
0 & \frac{4}{\alpha}
\end{array}
\right)\, ,
\end{equation}
where
\begin{equation}
\tilde G_{ij}=G^{eff}_{ij}-\frac{4}{\mathcal A}(Ba)_i(aB)_j\, ,
\end{equation}
and
\begin{equation}\label{eq:effm}
{}^\star G^{eff}_{ij}=G_{ij}^{eff}-\alpha a_i
a_j-\frac{4}{\mathcal A}(Ba)_i (aB)_j=\tilde G_{ij}-\alpha a_i
a_j\, .
\end{equation}
The corresponding inverse ones are of the form
\begin{equation}\label{eq:inveffex}
(G_{eff}^{-1})^{AB}=\left(
\begin{array}{cc}
(\tilde G^{-1})^{ij}+\frac{\tilde a^i \tilde a^j}{\tilde \mathcal A} & -\frac{\tilde a^i}{2\tilde \mathcal A}\\
-\frac{\tilde a^j}{2\tilde \mathcal A} & \frac{1}{4 \tilde
\mathcal A}
\end{array}\right)\, ,\quad ({}^\star G_{eff}^{-1})^{AB}=\left(
\begin{array}{cc}
({}^\star G_{eff}^{-1})^{ij} & 0\\
0 & \frac{\alpha}{4}
\end{array}
\right)\, ,
\end{equation}
with the space-time components
\begin{equation}\label{eq:relacije1}
(\tilde G^{-1})^{ij}=(G_{eff}^{-1})^{ij}+\frac{4}{\tilde \mathcal
A}(B\tilde a)^i (\tilde a B)^j\, ,\quad ({}^\star
G_{eff}^{-1})^{ij}=(G_{eff}^{-1})^{ij}+\frac{1}{\tilde \mathcal
A}\left[ \tilde a^i \tilde a^j+4(B\tilde a)^i(\tilde a
B)^j\right]\, ,
\end{equation}
and
\begin{equation}
\tilde \mathcal A\equiv \frac{1}{\alpha}-\tilde a^2\, .
\end{equation}
Because of the first relation in (\ref{eq:relacije1}), we can raise the
index of $a_i$ with both $(G_{eff}^{-1})^{ij}$ and $(\tilde
G^{-1})^{ij}$
\begin{equation}
\tilde a^i=(\tilde G^{-1})^{ij}a_j=(G_{eff}^{-1})^{ij}a_j\, ,\quad \tilde a^2=\tilde a^i a_i\, .
\end{equation}

The determinants of the effective metrics are
\begin{equation}
\det G_{AB}^{eff}=\det\; {}^\star G_{AB}^{eff}=4\frac{\tilde
\mathcal A^2}{\mathcal A}\det G_{ij}^{eff}\, .
\end{equation}

\subsection{Extended metrics in the zero central charge limit ($c=0$)}

In the zero central charge limit all quantities (including closed
and open string metric and projectors) in component form agree
with the expressions in the case without Liouville term (see
Ref.\cite{BNBS1})
\begin{equation}\label{eq:Gnula}
{}^0 G_{AB}=\left(
\begin{array}{cc}
G_{ij} & 2a_i\\
2a_j & 0
\end{array}\right)\, ,\quad {}^0 G^{eff}_{AB}=\left(
\begin{array}{cc}
{}^0 \tilde G_{ij} & 2a_i\\
2a_j & 0
\end{array}\right)\, ,
\end{equation}
while the inverse ones are
\begin{equation}\label{eq:inGnula}
({}^0 G^{-1})^{AB}=\left(
\begin{array}{cc}
(G^{-1}P_T^0)^{ij} & \frac{a^i}{2a^2}\\
\frac{a^j}{2a^2} & -\frac{1}{4a^2}
\end{array}\right)\, ,\quad ({}^0 G^{-1}_{eff})^{AB}=\left(
\begin{array}{cc}
\tilde P_T^{ij} & \frac{\tilde a^i}{2\tilde a^2}\\
\frac{\tilde a^j}{2\tilde a^2} & -\frac{1}{4\tilde a^2}
\end{array}\right)\, ,
\end{equation}
where
\begin{equation}\label{eq:tildeptij}
(P_T^0)_i{}^j=\delta_i{}^j-\frac{a_ia^j}{a^2}\, ,\quad \tilde P_T^{ij}=({}^0 \tilde G^{-1})^{ij}-\frac{\tilde a^i\tilde a^j}{\tilde a^2}\, .
\end{equation}
The term ${}^0 \tilde G_{ij}$ is defined as
\begin{equation}
{}^0 \tilde G_{ij}=G^{eff}_{ij}+\frac{4}{a^2}(Ba)_i(aB)_j\, ,
\end{equation}
while its inverse is
\begin{equation}
({}^0 \tilde G^{-1})^{ij}=(G^{-1}_{eff})^{ij}-\frac{4}{\tilde a^2}(B\tilde a)_i(\tilde aB)_j\, .
\end{equation}

\section{Extended space-time projectors}
\setcounter{equation}{0}

In this appendix we introduce projector operators in order to
separate noncommutative and nonphysical variables on the $Dp$-brane
as well as to express the noncommutativity parameter.

The projectors on the subspace spanned by vectors $n^A_a$, and on
the orthogonal one with respect to the metric $g_{AB}$ are
\begin{equation}\label{eq:definicija}
(\Pi)_A{}^B=\gamma^{ab}n_{a A} n_{b}^{B}\, ,\quad (\Pi_T)_A{}^B=\delta_A{}^B-(\Pi)_A{}^B\, ,
\end{equation}
where
\begin{equation}
n_{a A}=g_{AB} n_a^B\, ,\quad \gamma_{ab}=n_a^A\, g_{AB}\, n_b^B\, ,\quad \gamma^{ac}\gamma_{cb}=\delta^a{}_b\, .
\end{equation}
The transposed operator is
\begin{equation}\label{eq:trans}
\Pi^A{}_B=g^{AC}\Pi_C{}^D g_{DB}\, .
\end{equation}

\subsection{Projectors on regular part of $G_{AB}$}

In both representations the extended metrics, $G_{AB}$ and ${}^\star G_{AB}$, are singular
for $\mathcal A=0$. The corresponding singular directions are
\begin{equation}\label{eq:singG1}
n^A=\left(
\begin{array}{c}
a^i\\ -\frac{\alpha a^2}{2}
\end{array}
\right)\, ,\quad {}^\star n^A=M^A{}_B n^B=\left(
\begin{array}{c}
a^i\\ 0
\end{array}
\right)\, ,
\end{equation}
so that the vectors
\begin{equation}\label{eq:singG2}
n_A\equiv G_{AB}n^B=\alpha \mathcal A \left(
\begin{array}{c}
a_i\\0
\end{array}
\right)={}^\star n_A\equiv {}^\star G_{AB} {}^\star n^B\, .
\end{equation}
vanish for $\mathcal A=0$.

The corresponding projectors for non-diagonal and diagonal case
are
\begin{equation}
(P_T^0)_A{}^B=\left(\begin{array}{cc}
\delta_i{}^j-\frac{a_ia^j}{a^2} & \frac{\alpha a_i}{2}\\
0 & 1
\end{array}
\right)\, ,\quad ({}^\star P_T^0)_A{}^B=\left(\begin{array}{cc}
\delta_i{}^j-\frac{a_ia^j}{a^2} & 0\\
0 & 1
\end{array}
\right)\, .
\end{equation}
The metric in the subspace orthogonal to the vector $n^A$ is defined as
\begin{equation}
g_{AB}=(P_T^0 G)_{AB}\, ,\quad (g^{-1})^{AB}=(G^{-1}P_T^0)^{AB}\, .
\end{equation}
The concrete expressions are
\begin{equation}
g_{AB}=\left(\begin{array}{cc}
G_{ij}-\alpha \mathcal A\frac{a_i a_j}{a^2} & 2a_i\\
2a_j & \frac{4}{\alpha}
\end{array}
\right)\, ,\quad {}^\star g_{AB}=\left(\begin{array}{cc}
G_{ij}-\frac{a_i a_j}{a^2} & 0\\
0 & \frac{4}{\alpha}
\end{array}
\right)\, ,
\end{equation}
while the inverse ones have the same form in both representations
\begin{equation}\label{eq:malogmetrika}
(g^{-1})^{AB}=\left(\begin{array}{cc}
G^{ij}-\frac{a^i a^j}{a^2} & 0\\
0 & \frac{\alpha}{4}
\end{array}
\right)=({}^\star g^{-1})^{AB}\, .
\end{equation}

\subsection{Projectors on regular part of $G_{AB}^{eff}$}

There are two singular directions of the effective metric $G_{AB}^{eff}$
\begin{equation}\label{eq:vektilde1}
\tilde n^A_1=\left(
\begin{array}{c}
2(\tilde aB)^i\\0
\end{array}\right)\, ,\quad \tilde n^A_2=\left(
\begin{array}{c}
\tilde a^i\\-\frac{\alpha \tilde a^2}{2}
\end{array}\right)\, ,
\end{equation}
and of the corresponding one in diagonal representation ${}^\star G_{AB}^{eff}$ (\ref{eq:effmex})
\begin{equation}\label{eq:vektilde2}
{}^\star \tilde n^A_1=\left(
\begin{array}{c}
2(\tilde aB)^i\\0
\end{array}\right)\, ,\quad {}^\star \tilde n^A_2=\left(
\begin{array}{c}
\tilde a^i\\0
\end{array}\right)\, .
\end{equation}
In both representations the vectors with lower indices
\begin{equation}
\tilde n_{1 A}=\frac{\tilde \mathcal A}{\mathcal A}\left(
\begin{array}{c}
2(aB)_i \\ 0
\end{array}\right)={}^\star \tilde n_{1 A}\, ,\quad \tilde n_{2 A}=\alpha \tilde \mathcal A\left(
\begin{array}{c}
a_i \\ 0
\end{array}\right)={}^\star \tilde n_{2 A}\, ,
\end{equation}
vanish for $\tilde \mathcal A=0$.

Using the second expression (\ref{eq:definicija}) we obtain the
projectors on subspace orthogonal to the vectors $\tilde n_1^A$
and $\tilde n_2^A$
\begin{equation}\label{eq:hatpt}
(\hat P_T)_A{}^B=\left(
\begin{array}{cc}
(\hat P_T)_i{}^j & \frac{\alpha a_i}{2}\\
0 & 1
\end{array}\right)\, ,\quad ({}^\star \hat P_T)_A{}^B=\left(
\begin{array}{cc}
(\hat P_T)_i{}^j & 0\\
0 & 1
\end{array}\right)\, ,
\end{equation}
where $(\hat P_T)_i{}^j=\delta_i{}^j-\alpha a_i \tilde
a^j-\frac{4}{\mathcal A}(Ba)_i (\tilde a B)^j$.

\subsection{Metrics in subspace orthogonal on $\tilde n_1^A$ and $\tilde n_2^A$}

Using the definition of the extended effective metric
(\ref{eq:seffm}), we obtain for non-diagonal representation
\begin{equation}\label{eq:cekeff}
\check G_{AB}^{eff}=\left(
\begin{array}{cc}
\check G_{ij} & 2a_i\\
2a_j & \frac{4}{\alpha}
\end{array}\right)\, ,\quad (\check G_{eff}^{-1})^{AB}=\left(
\begin{array}{cc}
(\check G^{-1})^{ij}+\frac{\tilde a^i \tilde a^j}{\tilde \mathcal A} & -\frac{\tilde a^i}{2\tilde \mathcal A}\\
-\frac{\tilde a^j}{2\tilde \mathcal A} & \frac{1}{4\tilde \mathcal
A}
\end{array}\right)\, ,
\end{equation}
where
\begin{equation}
\check G_{ij}=G_{ij}^{eff}+\frac{4}{a^2}(Ba)_i (aB)_j\, ,\quad (\check G^{-1})^{ij}=(G_{eff})^{ij}-\frac{4}{\tilde a^2}(B\tilde a)^i(\tilde a B)^j\, .
\end{equation}
The corresponding expressions in diagonal form are obtained by
acting with similarity transformation (\ref{eq:slicnost})
\begin{equation}\label{eq:cekeffin}
{}^\star \check G_{AB}^{eff}=\left(
\begin{array}{cc}
\check G_{ij}-\alpha a_i a_j & 0\\
0 & \frac{4}{\alpha}
\end{array}\right)\, ,\quad ({}^\star \check G_{eff}^{-1})^{AB}=\left(
\begin{array}{cc}
(\check G^{-1})^{ij}+\frac{\tilde a^i \tilde a^j}{\tilde \mathcal A} & 0\\
0 & \frac{\alpha}{4}
\end{array}\right)\, ,
\end{equation}
Applying the procedure described at the beginning of this
appendix, on vectors $\tilde n_1^A$ and $\tilde n_2^A$ and on the
metric $\check G_{AB}^{eff}$, we obtain the projectors
\begin{equation}\label{eq:ptcek}
(\check P_T)_A{}^B=\left(
\begin{array}{cc}
(P_T)_i{}^j & \frac{\alpha a_i}{2}\\
0 & 1
\end{array}\right)\, ,\quad ({}^\star \check P_T)_A{}^B=\left(
\begin{array}{cc}
(P_T)_i{}^j & 0\\
0 & 1
\end{array}\right)\, ,
\end{equation}
for non-diagonal and diagonal case, respectively, where
$(P_T)_i{}^j=\delta_i{}^j-\frac{a_i \tilde a^j}{\tilde
a^2}+\frac{4}{\tilde \mathcal A}(Ba)_i (\tilde a B)^j$.

\end{document}